\documentclass[12pt]{article}

\newcommand{\be}{\begin{equation}}
\newcommand{\ee}{\end{equation}}
\newcommand{\bi}[1]{\vspace{-3mm} \bibitem{#1}}

\usepackage{epsfig,amsmath,amssymb,graphics,graphicx}

\begin{document}

\begin{center}
Physica A 387 (2008) 6505-6512
\vskip 5 mm
{\Large \bf Fokker-Planck equation with fractional coordinate derivatives }

\vskip 5 mm

{\large \bf Vasily E. Tarasov$^{1,2}$ and George M. Zaslavsky$^{1,3}$}\\

\vskip 3mm

{\it $1)$ Courant Institute of Mathematical Sciences, New York University \\
251 Mercer St., New York, NY 10012, USA, and }\\ 
{\it $2)$ Skobeltsyn Institute of Nuclear Physics, \\
Moscow State University, Moscow 119991, Russia } \\
{\it $3)$ Department of Physics, New York University, \\
2-4 Washington Place, New York, NY 10003, USA } \\

\vskip 11 mm

\begin{abstract}
Using the generalized Kolmogorov-Feller equation with long-range interaction,
we obtain  kinetic equations with fractional derivatives with respect to coordinates.  
The method of successive approximations 
with the averaging with respect to a fast variable is used. 
The main assumption is that the correlation function of 
probability densities of particles to make a step has a power-law dependence.
As a result, we obtain Fokker-Planck equation with fractional 
coordinate derivative of order $1<\alpha<2$.

\end{abstract}

\end{center}

\noindent
%%%PACS:  03.50.De; 05.45.Df; 41.20.-q  \\

\newpage
\section{Introduction}

In studying processes with fractal time and long-term memory 
a generalized kinetic equation was proposed in \cite{MontrollShlesinger}.
While the equation was of the master equation type, its  main property
was the presence of the power-type kernel for a probability density
to make a step. This type of equation was compared to the 
Kolmogorov-Feller equation in \cite{SZ}.
In this paper, we would like to go farther and to show the conditions under
which one can obtain a fractional generalization of the Fokker-Planck equation 
from the Kolmogorov-Feller equation.

Fractional calculus \cite{SKM,KST,Podlubny} has found many applications 
in recent studies in mechanics and physics, and the interest 
in fractional equations has been growing continually during the last years 
\cite{Mainardi,Hilfer,Zaslavsky1,Zaslavsky2,SATM,Agrawal,Baleanu}. 
Fractional Fokker-Planck equations with coordinate and time derivatives of 
non-integer order has been suggested in \cite{Zaslavsky7}. 
The solutions and properties of these equations 
are described in Refs. \cite{SZ,Zaslavsky1}.
The Fokker-Planck equation with fractional coordinate derivatives
was also considered in Refs. \cite{1994,Mil,Y,Chaos2005,Kol}.

The Kolmogorov-Feller equation is an integro-differential one? 
and it belongs to the type of master equations broadly used 
in different physical applications. 
It is well-known that the Kolmogorov-Feller equation can lead us 
to the Fokker-Planck equation \cite{Gnedenko}, under some conditions. 
In this paper, we use the method of successive approximations 
with averaging with respect to a fast variable \cite{Chaos}. 
We suppose that the correlation function of probability densities, 
which are used in the Kolmogorov-Feller equation, is a power type.
The Fokker-Planck equations with coordinate  
derivatives of order $1<\alpha<2$ are derived.

In Section 2, the Kolmogorov-Feller equation for 
the one-dimensional case is considered to 
fix notations and provide convenient references.
We note that power-law probability to make a step 
gives the equation with a fractional derivative.
In Section 3, we present a generalization of 
the Kolmogorov-Feller equation for the two-dimensional case.
The method of successive approximations is used for 
this generalized equation in Section 4.
In Section 5, we use averaging with respect to the fast variable 
to derive fractional Fokker-Planck equations.
Finally, a short conclusion is given in Section 6.

%%%%%%%%%%%%%%%%%%%%%%%%%%%%%%%%%%%%%%%%%%%%%%%

\section{Kolmogorov-Feller equation for one-dimensional case}

\subsection{Operator representation of the KF-equation}

Let $P(t,x)$ be a probability density to find a particle at $x$ at time instant $t$.
The normalization condition for $P(t,x)$ is 
\[ \int^{+\infty}_{-\infty} dx \, P(t,x) =1 \quad (t>0). \]
The Kolmogorov-Feller (KF) equation has the form
\be \label{KF1}
\frac{\partial P(t,x)}{\partial t}= 
\int^{+\infty}_{-\infty} dx' \, w(x') [P(t,x-x')-P(t,x)] , \quad P(0,x)=\delta(x) , \ee
where $w(x')$ is the probability density of particle to make a step
of the length $x'$, and
\be \label{nc} \int^{+\infty}_{-\infty} dx' \, w(x')=1 . \ee

Let us introduce the operator representation of the KF-equation.
We define the translation operator 
\be \label{TO1} T_{x'}=\exp \{ -x' \partial_x \} , \ee
such that
\be \label{TO2} T_{x'} P(t,x)=P(t,x-x') , \ee
and the finite difference operator 
\be \label{Delta-x} \Delta_{x'}=I-T_{x'} , \ee
where $I$ is an identity operator.
Then the Kolmogorov-Feller equation (\ref{KF1}) can be presented as
\be \label{KF2}
\frac{\partial P(t,x)}{\partial t}= L(\Delta) \, P(t,x) . \ee
Here we use the integro-differential operator
\be \label{KFO} L(\Delta)=-
\int^{+\infty}_{-\infty} dx' \, w(x') \, \Delta_{x'} . \ee
The operator (\ref{KFO}) will be called the Kolmogorov-Feller operator.
Note that power-law probability $w(x')$ in Eq. (\ref{KF1})
allows us to introduce a derivative of non-integer order \cite{KST}.

\subsection{KF-equation with fractional coordinate derivative}

The well-known fractional Caputo derivative \cite{KST} of order $\alpha$ is defined by
\be \label{FD} ^CD^{\alpha}_x P(x)=
\frac{1}{\Gamma(1-\alpha)} \int^x_{-\infty} \frac{d z}{(x-z)^{\alpha}}
\frac{\partial P(z)}{\partial z} , \quad (0<\alpha<1) . \ee
The fractional Marchaud derivative \cite{KST} of order $\alpha$ is defined by
\be \label{RFD}
{\bf D}^{\alpha}_x  P(x)=
\frac{1}{\Gamma(-\alpha)} \int^x_{-\infty} [P(z)-P(x)] \frac{d z}{(x-z)^{\alpha+1}} , 
\quad (0< \alpha <1). \ee
Using $x'=x-z$, equation (\ref{RFD}) has the form
\[ {\bf D}^{\alpha}_x P(x)= \frac{1}{\Gamma(-\alpha)} 
\int^{\infty}_{0} \frac{d x'}{(x')^{\alpha+1}} [P(x-x')-P(x)] . \]
If the function $w(x')$ in the KF-equation (\ref{KF1}) is the exponential function
up to a small parameter $\varepsilon$ such that
\be \label{wH} w(x')=\frac{a}{x'^{\alpha+1}} \, H(x') + O(\varepsilon) , \ee
where $H(x')$ is a Heaviside step function,
then Eq. (\ref{KF1}) can be presented as
%%%through the fractional coordinate derivative
\be \label{KF4}
\frac{\partial P(t,x)}{\partial t}= a {\bf D}^{\alpha}_x P(t,x) + O(\varepsilon) , 
\quad P(0,x)=\delta(x), \quad (0< \alpha <1). \ee
This equation has a fractional coordinate derivative
of order $0< \alpha <1$.
Note that the function $w(x)$ is a probability density? and it should satisfy 
the normalization condition (\ref{nc}).

\subsection{Generalized KF-equation}

In the general case, we can suppose that probability density of 
a particle to make a step of the length $x'$ depends 
on the time instant $t$ and the coordinate $x$. 
Then we should replace $w(x')$ by $w(t,x|x')$ in KF-equation (\ref{KF1}).
As a result, we can consider the equation
\be \label{KF1b-xy0}
\varepsilon^{-1} \partial_t P(t,x) = \varepsilon 
\int^{+\infty}_{-\infty} dx' \, w(t,x|x') [P(t,x-x')-P(t,x)] , 
\quad P(0,x)=\delta(x) , \ee
where $\varepsilon$ is a small parameter and $\partial_t=\partial / \partial t$. 
Note that the interpretation of the appearence of the small parameter $\varepsilon$ 
is connected with the change of the scale $t \to  \varepsilon t$,
such that  $\partial_t \, \to \, \varepsilon \, \partial_t$.
The function $w(t,x|x')$ is the probability density to make a step
of the length $x'$ at the time instant $t$ from the coordinate $x$. 
If $w(t,x|x')=w(x')$, then Eq. (\ref{KF1b-xy0}) gives Eq. (\ref{KF1}). 
We can assume that during any interval of time $(t,t+dt)$ the value of 
the variable $x(t)$ remains equal to $x$ with probability $1-p(t,x) dt$
and may undergo a change only with probability $p(t,x) dt$
(see more about this equation in Sec.55. of \cite{Gnedenko}). 
We suppose that 
\be \label{KF1b-xy}
w(t,x|x')=p(t,x) w(x') , \ee
where $p(t,x)$ is a bounded function.
If $p(t,x)=1$, then Eqs. (\ref{KF1b-xy0}) 
and (\ref{KF1b-xy}) gives Eq. (\ref{KF1}).

Using the operator (\ref{Delta-x}),
equation (\ref{KF1b-xy0}) can be presented in the operator form
\be \label{KF2b}
\frac{\partial P(t,x)}{\partial t}= \varepsilon L(t,x,\Delta) \, P(t,x) , \ee
where 
\be \label{KFOb} L(t,x,\Delta)=-
\int^{+\infty}_{-\infty} dx' \, w(t,x|x') \, \Delta_{x'} . \ee
Equation (\ref{KF2b}) will be called a generalized KF-equation
for one-dimensional case.

%%%%%%%%%%%%%%%%%%%%%%%%%%%%%%%%%%%%%%%%%%%%%%%%%%%%%%%%%%%%%%%%%%%%%%%%%
\subsection{Successive approximations}

The generalized Kolmogorov-Feller equation (\ref{KF2b}) 
can be rewritten in the integral form
\be \label{KF3} P(t,x)-P(0,x)= 
\varepsilon \int^t_0 d\tau \ L(\tau,x,\Delta) P(\tau,x) . \ee
This equation can be presented as the integral Volterra type equation
\be \label{41}
P(t,x)=P(0,x)+ \varepsilon \int^{t}_{0} dt_1 \,
L(t_1,x,\Delta) P(t_1,x) .  \ee
%%%Changing the variable $t_1 \rightarrow t_2$, and 
%%%$t \rightarrow t_1$ in equation (\ref{4}).
Let us consider the successive approximations.
Substitution of equation (\ref{41}) in the form
\be \label{51}
P(t_1,x)=P(0,x)+ \varepsilon \int^{t_1}_{0} dt_2 \, L(t_2,x\Delta) P(t_2,x) \ee
into equation (\ref{41}) gives
\[ P(t,x)=P(t_0,x)+ \varepsilon \int^{t}_{0} dt_1 \, L(t_1,x,\Delta) P(0,x) +  \]
\be \label{52}
+ \varepsilon^2 \int^{t}_{0} dt_1 \int^{t_1}_{0} dt_2 \,
L(t_1,x,\Delta) L(t_2,x,\Delta) P(t_2,x) .
\ee
Changing the variables $t_1 \rightarrow t_2$, and $t \rightarrow t_1$ 
in equation (\ref{51}), and substituting into (\ref{52}), we obtain
an equation up to $\varepsilon^2 $ in the form  
\[ P(t,x)=P(0,x)+ \varepsilon \int^{t}_{0} dt_1 \, 
L(t_1,x,\Delta) P(0,x) +  \]
\be \label{53}
+ \varepsilon^2 \int^{t}_{0} dt_1 \int^{t_1}_{0} dt_2 \, 
L(t_1,x,\Delta) L(t_2,x,\Delta) P(0,x)  .
\ee
If the function $w(x')$ is the exponential function (\ref{wH}), 
then $L(t,x,\Delta)$ is a differential operator of order $0<\alpha<1$
with respect to $x$, and 
$L(t_1,x,\Delta) L(t_2,x,\Delta) P(t_2,x)$ is a differential operator 
of the order $0<2 \alpha<2$.
To obtain fractional kinetic equations of the order $0<2 \alpha<2$, 
we should consider the averaging procedure before a 
partial differentiation of Eq. (\ref{53}) with respect to time
is realized. 
Without averaging, we derive an equation of order $0<\alpha<1$.

%%%%%%%%%%%%%%%%%%%%%%%%%%%%%%%%%%%%%%%%%%%%%%%%%%%%%%%%%%%%%%%%%%%%%%%%%%
\section{Distribution function and Kolmogorov-Feller equation for two-dimensional case}

Let $P(t,x,y)$ be a function of probability distribution in a phase space.
The variables $x$ and $y$ describe the phase space of a system.
There are the following interpretations for the variables $x$ and $y$.

\begin{itemize}
\item A system with one degree of freedom can be presented by
momentum $x=p$, and coordinate $y=q$. 

\item A system can be described by action $x=I$, and phase $y=\theta$.

\item $n$-particle system can be defined by $x=(q_1,p_1)$ and
$y=(q_2,p_2,...,q_n,p_n)$,
or $x=q_1$, and $ y=(p_1,q_2,p_2,...,q_n,p_n )$. 

\item The variables $x$ describe a system, and $y$ describes
an environment of this system.
\end{itemize}

We plan to use the reduced distribution and 
average values with respect to $y$, 
where $y$ will be considered as a fast variable.

\subsection{Generalized KF-equation for two-dimensional case}

We assume that $x\in X \subset \mathbb{R}$ and $y\in Y \subset \mathbb{R}$,
then ${\bf r}=(x,y) \in X \times Y \subset \mathbb{R}^2$. 
We plan to consider $x$ as a slow variable 
and $y$ will be considered as a fast variable. 
The distribution function $P(t,{\bf r})$ in the region $X \times Y$ 
satisfies the generalized Kolmogorov-Feller equation
\be \label{KF1b-r}
\frac{\partial P(t,{\bf r})}{\partial t}= \varepsilon
\int_{X \times Y} d^2 r_1 \, w(t,{\bf r}|{\bf r}_1) 
[P(t,{\bf r}-{\bf r}_1)-P(t,{\bf r})] , \ee
where $d^2 r_1=dx_1 dy_1$, and $\varepsilon$ is a small parameter. 
Here $w(t,{\bf r}|{\bf r}_1)$ is the probability density to make a step
${\bf r}_1$ at the time instant $t$ from the point ${\bf r}$. 

Equation (\ref{KF1b-r}) can be presented in the form
\be \label{1r} \frac{\partial}{\partial t} P(t,{\bf r})
=\varepsilon L(t,{\bf r},\Delta) P(t,{\bf r}) , \ee
where $L(t,{\bf r},\Delta)$ is a Kolmogorov-Feller operator
\be \label{KFO-2aa} L(t,{\bf r},\Delta) =
-\int_{X \times Y} d^2 r_1 \, w(t,{\bf r} | {\bf r}_1) \, \Delta_{{\bf r}_1} . \ee
Here $\Delta_{{\bf r}_1}$ is a finite difference operator
\[ \Delta_{{\bf r}_1}=I - T_{{\bf r}_1} , \] 
where $T_{{\bf r}_1}$ is a translation operator in $X \times Y$, that is defined by
\[ T_{{\bf r}_1}= \exp \{- {\bf r}_1 \nabla\}=
\exp\{-x_1 \partial_x -y_1 \partial_y \} . \]
We can assume that during any interval of time $(t,t+dt)$ the value of 
the variable ${\bf r}(t)$ remains equal to ${\bf r}$ 
with probability $1-p(t,{\bf r}) dt$
and may undergo a change only with probability $p(t,{\bf r}) dt$
(see Sec. 55. in \cite{Gnedenko}). Then
\be 
w(t,{\bf  r}|{\bf r}')=p(t,{\bf r}) \, w({\bf r}') , \ee
where $p(t,{\bf r})$ is a bounded function.

We can use the variables $x$, $y$ instead of ${\bf r}$.
Then equation (\ref{KF1b-r}) for the distribution function $P(t,x,y)$ 
will be presented in the form
\be \label{KF1b-2a}
\frac{\partial P(t,x,y)}{\partial t}= \varepsilon
\int_{X \times Y} dx_1 dy_1 \, w(t,x,y|x_1,y_1) [P(t,x-x_1,y-y_1)-P(t,x,y)] . \ee
%%%where $\varepsilon$ is a small parameter.
This equation can be rewritten as
\be \label{1} \frac{\partial}{\partial t} P(t,x,y)=
\varepsilon L(t,x,y,\Delta_x,\Delta_y) P(t,x,y) , \ee
where $L(t,x,y,\Delta_x,\Delta_y)$ is a Kolmogorov-Feller operator
\be \label{KFO-2a} L(t,x,y,\Delta_x,\Delta_y)=
\int_{X \times Y} dx_1 dy_1 \, w(t,x,y|x',y') \, [T_{x_1}T_{y_1}-I] . \ee
This is the generalized Kolmogorov-Feller equation in the operator form.

%%%%%%%%%%%%%%%%%%%%%%%%%%%%%%%%%%%%%%%%%%%%%%%%
\section{Method of successive approximations}

The generalized Kolmogorov-Feller equation 
\be \label{3}
\frac{\partial}{\partial t}P(t,{\bf r})=\varepsilon L(t,{\bf r},\Delta) P(t,{\bf r})  \ee
can be presented as the integral Volterra equation
\be \label{4}
P(t,{\bf r})=P_0({\bf r})+ \varepsilon \int^{t}_{0} dt_1 \,
L(t_1,{\bf r},\Delta) P(t_1,{\bf r}) ,  \ee
where $P_0({\bf r})= P(0,{\bf r})$. 
Substitution of equation (\ref{4}) in the form
\be \label{5}
P(t_1,{\bf r})=P_0({\bf r})+ \varepsilon \int^{t_1}_{0} dt_2 \,
L(t_2,{\bf r},\Delta) P(t_2,{\bf r})  \ee
into Eq. (\ref{4}) gives
\[ P(t,{\bf r})=P_0({\bf r})+ \varepsilon \int^{t}_{0} dt_1 \, 
L(t_1,{\bf r},\Delta) P_0({\bf r}) +  \]
\be \label{5b}
+\varepsilon^2 \int^{t}_{0} dt_1 \int^{t_1}_{0} dt_2 \,
L(t_1,{\bf r},\Delta) L(t_2,{\bf r},\Delta) P(t_2,{\bf r}) .
\ee
Changing the variables $t_1 \rightarrow t_2$, and $t \rightarrow t_1$ 
in equation (\ref{5}), and substituting into (\ref{5b}), we get
\[ P(t,{\bf r})=P_0({\bf r})+ \varepsilon \int^{t}_{0} dt_1 \, 
L(t_1,{\bf r},\Delta) P_0({\bf r}) +  \]
\be \label{5c}
+ \varepsilon^2 \int^{t}_{0} dt_1 \int^{t_1}_{0} dt_2 \, 
L(t_1,{\bf r},\Delta) L(t_2,{\bf r},\Delta) P_0({\bf r}) +...
\ee
Using the chronological multiplication
\be
T \{ L(t_1,{\bf r},\Delta) L(t_2,{\bf r},\Delta)   \}=
\left\{
\begin{array}{ll}
L(t_1,{\bf r},\Delta) L(t_2,{\bf r},\Delta) &  t_1 >t_2 ;\\
L(t_2,{\bf r},\Delta) L(t_1,{\bf r},\Delta) & t_2 >t_1 ,
\end{array} \right.
\ee
equation (\ref{5c}) can be symmetric with respect to $t_1$ and $t_2$: 
\[ P(t,{\bf r})=P_0({\bf r})+ \varepsilon \int^{t}_{0}
dt_1 \, L(t_1,{\bf r},\Delta) P_0({\bf r}) +  \]
\be \label{6}
+ \frac{1}{2}\varepsilon^2 \int^{t}_{0} dt_1 \int^{t}_{0}
dt_2 \, T\{ L(t_1,{\bf r},\Delta) L(t_2,{\bf r},\Delta) \} P_0 ({\bf r}) +...
\ee
This is the symmetric representation of equation (\ref{5c}).

%%%%%%%%%%%%%%%%%%%%%%%%%%%%%%%%%%%%%%%%%%%%%%%%%%%%%%%%%%%

\section{Averaging with respect to the fast variable}

In this section, let us consider the variables ${\bf r}=(x,y)$ 
as slow ($x$) and fast ($y$).
Substitution of (\ref{KFO-2aa}) into (\ref{6}) gives
\[ P(t,{\bf r})=P_0({\bf r})- \varepsilon \int^{t}_{0} dt_1 \, \int_{X \times Y} 
d^2 r_1 \, w(t_1,{\bf r} | {\bf r}_1) \, \Delta_{{\bf r}_1} P_0({\bf r}) +  \]
\be \label{6-2}
+ \frac{1}{2}\varepsilon^2 \int^{t}_{0} dt_1 \int^{t}_{0} dt_2 \, 
\int_{X \times Y} d^2 r_1 \, \int_{X \times Y} d^2 r_2 \, 
T\{w(t_1,{\bf r} | {\bf r}_1) \, \Delta_{{\bf r}_1} \, 
 w(t_2,{\bf r} | {\bf r}_2) \, \Delta_{{\bf r}_2} \}
P_0 ({\bf r}) + O(\varepsilon^3) .
\ee

The function $w(t,{\bf r}|{\bf r}_1)$ is the probability density to make a step
of the vector ${\bf r}_1$ at the time instant $t$ from the point ${\bf r}$. 
The first assumption is that this probability density has a weak dependence 
(up to terms of order $O(\varepsilon)$) on the point ${\bf r}$, i.e.,
\be w(t,{\bf r}+{\bf r}_1 | {\bf r}_2) =w(t,{\bf r} | {\bf r}_2) + O(\varepsilon) . \ee
Then 
\[ \Delta_{{\bf r}_1} \,  w(t_2,{\bf r} | {\bf r}_2)=0+ O(\varepsilon)  , \]
and
\[ w(t_1,{\bf r} | {\bf r}_1) \, \Delta_{{\bf r}_1} \, 
w(t_2,{\bf r} | {\bf r}_2) \, \Delta_{{\bf r}_2} =
w(t_1,{\bf r} | {\bf r}_1) \,  w(t_2,{\bf r} | {\bf r}_2) 
\, \Delta_{{\bf r}_1} \, \Delta_{{\bf r}_2}+ O(\varepsilon) . \]
As a result, equation (\ref{6-2}) has the form
\[ P(t,{\bf r})=P_0({\bf r})- \varepsilon \int^{t}_{0}
dt_1 \, \int_{X \times Y} d^2 r_1 \, w(t_1,{\bf r} | {\bf r}_1) \, \Delta_{{\bf r}_1} 
P_0({\bf r}) +  \]
\[ + \frac{1}{2}\varepsilon^2 \int^{t}_{0} dt_1 \int^{t}_{0} dt_2 \, 
\int_{X \times Y} d^2 r_1 \, \int_{X \times Y} d^2 r_2 \, 
T \{w(t_1,{\bf r} | {\bf r}_1) \,  w(t_2,{\bf r} | {\bf r}_2) \} 
\, \Delta_{{\bf r}_1} \, \Delta_{{\bf r}_2} \, P_0 ({\bf r}) + O(\varepsilon^3) . \]

The second assumption states that $P_0 ({\bf r})=P_0 (x,y)$ 
does not depend on the fast variable $y$ up to $ \varepsilon $-terms, such that
\be P_0 ({\bf r}) =P_0 (x,y) =\rho_0(x) + O(\varepsilon) . \ee
Then, we have
\be
\Delta_{{\bf r}_1} \, \Delta_{{\bf r}_2} \, P_0 ({\bf r}) =
\Delta_{x_1} \, \Delta_{x_2} \, \rho_0 (x) + O(\varepsilon) .
\ee

Averaging over the variable $y$ will be denoted by $ < \ >_y $.
We also use the notations
\be
\rho(t,x)=<P(t,{\bf r})>_y ,
\ee
and $\rho(0,x)=\rho_0(x)$.

Averaging of equation (\ref{6-2}) with respect to the fast variable $y$, we obtain
\[ \rho(t,x)=\rho_0(x)- \varepsilon \int^{t}_{0} dt_1 \, \int_{X} d x_1 \, 
A(t_1,x| x_1) \, \Delta_{x_1} \rho_0(x) +  \]
\begin{equation} \label{6-4}
+ \frac{1}{2}\varepsilon^2 \int^{t}_{0} dt_1 \, 
\int_{X} d x_1 \, \int_{X} d x_2 \, B(t_1,x|x_1,x_2) \,   
\Delta_{x_1} \, \Delta_{x_2} \, \rho_0 (x) + O(\varepsilon^3) ,
\end{equation}
where we introduce the functions 
\begin{equation}
A(t_1,x| x_1)= \int_{Y} d y_1 \, < w(t_1, {\bf r} | {\bf r}_1) >_y ,
\end{equation}
\begin{equation}
B(t_1,x|x_1,x_2) = \int^{t}_{0} dt_2 \, \int_{Y} d y_1 \, \int_{Y} d y_2 \, 
<T \{w(t_1,{\bf r} | {\bf r}_1) \,  w(t_2,{\bf r} | {\bf r}_2) \}>_y .
\end{equation}
Using ${\bf r}=(x,y)$, these functions can be presented by
\begin{equation}
A(t_1,x| x_1)= \int_{Y} d y_1 \, < w(t_1,x,y | x_1, y_1) >_y ,
\end{equation}
\begin{equation}
B(t_1,x|x_1,x_2) = \int^{t}_{0} dt_2 \, \int_{Y} d y_1 \, \int_{Y} d y_2 \, 
<T \{w(t_1,x,y| x_1,y_1) \,  w(t_2,x,y | x_2,y_2) \}>_y .
\end{equation}

The third assumption is that the function $B(t_1,x|x_1,x_2)$ is diagonal 
with respect to variables $x_1$ and $x_2$ up to $\varepsilon$-term, i.e.,
\begin{equation}
B(t_1,x|x_1,x_2) = B(t_1,x|x_1) \delta (x_1-x_2)+ O(\varepsilon) .
\end{equation}
Then the operator $\Delta_{x_1} \, \Delta_{x_2}$ 
will be the finite difference operator $\Delta^2_{x_1}$ of second order.
This allows us to have fractional derivative for 
the exponential function $B(t,x|x_1)$ since 
Marchaud and Riesz fractional derivatives \cite{SKM} are defined through 
the finite difference operator.

The fourth assumption is that the functions $A(t_1,x|x_1)$ and $B(t_1,x|x_1)$
are exponential functions up to $\varepsilon$ such that
\be \label{A1}
A(t,x|x_1)=a(t) \frac{1}{\kappa(\alpha_1,1)} 
\frac{1}{|x_1|^{\alpha_1+1}} H(x_1)+ O(\varepsilon) , \quad (0<\alpha_1<1) ,
\ee
\be \label{B1}
B(t,x|x_1)=b(t,x) \frac{1}{\kappa(\alpha_2,2)} 
\frac{1}{|x_1|^{\alpha_2}} H(x_1)+ O(\varepsilon) , \quad (1<\alpha_2<2) ,
\ee
where $H(x)$ is the Heaviside step function, and
\be
\kappa (\alpha,n)=-\Gamma(\alpha) A_n(\alpha) , \quad
A_n(\alpha)=\sum^n_{k=0} (-1)^{k-1} \frac{n!}{k! (n-k)!} k^{\alpha} , 
\ee
with $n-1<\alpha<n$. 

As a result, Eq. (\ref{6-4}) gives
\be \label{rho-1}
\rho(t,x)=\rho_0(x)+ \varepsilon \int^t_{0} dt_1 \, a(t_1) {\bf D}^{\alpha_1}_x \rho_0(x)
+ \varepsilon^2 \int^t_{0} dt_1 \, b(t_1,x) {\bf D}^{\alpha_2}_x \rho_0(x) 
+ O(\varepsilon^3) .
\ee
Here ${\bf D}^{\alpha}_x$ is Marchaud fractional derivative \cite{SKM}
of order $\alpha$ with respect to $x$, which is defined by the equation
\be
{\bf D}^{\alpha}_x f(x)
=\frac{1}{\kappa (\alpha,n)} \int^{\infty}_0 \frac{\Delta^n_{y} f(x)}{y^{\alpha+2-n}} dy, 
\quad (n-1<\alpha<n) .
\ee
where $\Delta^n_{y}$ is a finite difference of order $n$ such that
\be
\Delta^n_{y} f(x) = (I-T_{y})^n f(x)=
\sum^n_{m=0} (-1)^m \frac{n!}{m! (n-m)!} f(x-my) .
\ee

In general, the variables $x$ and $x_1$ can be vectors 
in the $N$-dimensional space $\mathbb{R}^N$, where $N=1,2,3,...$. 
The fourth assumption for the functions $A(t_1,x|x_1)$ and $B(t_1,x|x_1)$
can be realized in the form other than (\ref{A1}) and (\ref{B1}).
We can suppose that the functions $A(t_1,x|x_1)$ and $B(t_1,x|x_1)$
are exponential functions up to $\varepsilon$ such that
\be \label{A2}
A(t,x|x_1)=a(t) \frac{1}{d_N(1,\alpha_1)} 
\frac{1}{|x_1|^{N+\alpha_1}} + O(\varepsilon) , \quad (0<\alpha_1<1) ,
\ee
\be \label{B2}
B(t,x|x_1)=b(t,x) \frac{1}{d_N(2,\alpha_2)} 
\frac{1}{|x_1|^{N+\alpha_2}}+ O(\varepsilon) , \quad (1<\alpha_2<2) ,
\ee
where $x,x_1 \in \mathbb{R}^N$, and
\be
d_N (n,\alpha)=\frac{2^{-\alpha} \pi^{1+N/2} A_n(\alpha)}{ \sin(\alpha \pi /2) 
\Gamma(1+\alpha/2) \Gamma((N+\alpha)/2) }
%%%A_n(\alpha)=\sum^n_{k=0} (-1)^{k-1} \frac{n!}{k! (n-k)!} k^{\alpha} , 
\ee
with $n-1<\alpha<n$. 
As a result, Eq. (\ref{6-4}) gives a fractional equation of the form (\ref{rho-1}), 
where ${\bf D}^{\alpha}_x$ is the fractional Riesz derivative (see \cite{SKM} Sec.25.4)
of order $\alpha$ with respect to $x \in \mathbb{R}^N$, 
which is defined by the equation
\be \label{RD}
{\bf D}^{\alpha}_x f(x)
=\frac{1}{d_N(n,\alpha)} \int_{\mathbb{R}^N} 
\frac{\Delta^n_{y} f(x)}{|y|^{\alpha+N}} \, d^Ny, 
\quad (n-1<\alpha<n) .
\ee
To denote the Riesz fractional derivative (\ref{RD}),  
the notation $(-\Delta)^{\alpha/2}$ is also used. 
Note that the Fourier transform ${\cal F}$ of this derivative 
(see Property 2.34 of \cite{KST})
is defined by
\[ \Bigl({\cal F}\{ {\bf D}^{\alpha}_x f(x) \} \Bigr)({\bf k}) =
|{\bf k}|^{\alpha} \, \Bigl( \{ {\cal F} f(x) \} \Bigr) ({\bf k}) .  \] 
The representation of the assumption in the form (\ref{A2}) and (\ref{B2})
allows us to obtain a fractional kinetic equation 
for arbitrary $N$-dimensional space
(for example, in the 3-dimensional Euclidean space).

The partial time differentiation of equation (\ref{rho-1}) gives
\be \label{rho-2}
\frac{\partial}{\partial t}\rho(t,x)=\varepsilon a(t) {\bf D}^{\alpha_1}_x \rho_0(x)
+ \varepsilon^2  b(t,x) {\bf D}^{\alpha_2}_x \rho_0(x) + O(\varepsilon^3) .
\ee
Substitution of equation (\ref{rho-1}) in the form
\be
\rho_0(x)=\rho(t,x)- 
\varepsilon \int^t_{0} dt_1 \, a(t_1) {\bf D}^{\alpha_1}_x \rho_0(x)
+ O(\varepsilon^2)
\ee
into equation (\ref{rho-2}) gives
\be \label{rho-3}
\frac{\partial}{\partial t}\rho(t,x)=\varepsilon a(t) {\bf D}^{\alpha_1}_x \rho(t,x)
+ \varepsilon^2 \Bigl( b(t,x) {\bf D}^{\alpha_2}_x -
c(t) {\bf D}^{2\alpha_1}_x \Bigr) \rho(t,x) + O(\varepsilon^3) ,
\ee
where
\[ c(t)=a(t) \int^t_0 dt_1 \, a(t_1) . \]

Equation (\ref{rho-3}) up to $ O(\varepsilon^3)$ has the form
\be \label{rho-4}
\frac{\partial}{\partial t}\rho(t,x)=\varepsilon a(t) {\bf D}^{\alpha_1}_x \rho(t,x)
+ \varepsilon^2 \Bigl( b(t,x) {\bf D}^{\alpha_2}_x -
c(t){\bf D}^{2\alpha_1}_x \Bigr) \rho(t,x) .
\ee
This is the fractional kinetic equation with non-integer derivatives of the order
$1<\alpha_2<2$ and $0<2\alpha_1<2$ with respect to coordinate $x$.

If $a(t)=0$, i.e. $A(t,x|x_1)=0$, then 
we have the fractional equation of order $1<\alpha_2<2$ 
with respect to $x$, such that
\be \label{rho-5}
\frac{\partial}{\partial t}\rho(t,x)= 
\varepsilon^2 b(t,x) {\bf D}^{\alpha_2}_x \rho(t,x) .
\ee
This is the fractional Fokker-Planck equation that 
is suggested in \cite{Zaslavsky7} to describe fractional kinetics.

%%%%%%%%%%%%%%%%%%%%%%%%%%%%%%%%%%%%%%%%%%%%%%%%%%%%%%%%%%%%%%%%%%%%%%%%%%

\section{Conclusion}

In this paper, we consider the Fokker-Planck equations 
with coordinate derivatives of non-integer order $1<\alpha<2$.
These derivatives are defined in the form of the fractional 
Marchaud and Riesz differentiations.
The starting point of our consideration is 
the well-known Kolmogorov-Feller equation, 
and some generalization of the equation.
The fractional kinetic equations are derived by using
the method of successive approximations, 
and the averaging with respect to a fast variable. 
In the paper, we assume that the correlation function 
of probability density for the generalized Kolmogorov-Feller equation 
has power-law form.
Note that some properties the Fokker-Planck equation with 
coordinate derivatives of non-integer order $1<\alpha<2$  
are considered in \cite{SZ,Zaslavsky1,1994,Mil,Y,Chaos2005,Kol}.

%%%%%%%%%%%%%%%%%%%%%%%%%%%%%%%%%%%%%%%%%%%%%%

\end{document}